\documentstyle[12pt]{article}
\newread\epsffilein    % file to \read
\newif\ifepsffileok    % continue looking for the bounding box?
\newif\ifepsfbbfound   % success?
\newif\ifepsfverbose   % report what you're making?
\newdimen\epsfxsize    % horizontal size after scaling
\newdimen\epsfysize    % vertical size after scaling
\newdimen\epsftsize    % horizontal size before scaling
\newdimen\epsfrsize    % vertical size before scaling
\newdimen\epsftmp      % register for arithmetic manipulation
\newdimen\pspoints     % conversion factor
\def\epsfbox#1{\global\def\epsfllx{72}\global\def\epsflly{72}%
   \global\def\epsfurx{540}\global\def\epsfury{720}%
   \def\lbracket{[}\def\testit{#1}\ifx\testit\lbracket
   \let\next=\epsfgetlitbb\else\let\next=\epsfnormal\fi\next{#1}}%
\def\epsfgetlitbb#1#2 #3 #4 #5]#6{\epsfgrab #2 #3 #4 #5 .\\%
   \epsfsetgraph{#6}}%
\def\epsfnormal#1{\epsfgetbb{#1}\epsfsetgraph{#1}}%
\def\epsfgetbb#1{%
\openin\epsffilein=#1
\ifeof\epsffilein\errmessage{I couldn't open #1, will ignore it}\else
   {\epsffileoktrue \chardef\other=12
    \def\do##1{\catcode`##1=\other}\dospecials \catcode`\ =10
    \loop
       \read\epsffilein to \epsffileline
       \ifeof\epsffilein\epsffileokfalse\else
          \expandafter\epsfaux\epsffileline:. \\%
       \fi
   \ifepsffileok\repeat
   \ifepsfbbfound\else
    \ifepsfverbose\message{No bounding box comment in #1; using defaults}\fi\fi
   }\closein\epsffilein\fi}%
\def\epsfsetgraph#1{%
   \epsfrsize=\epsfury\pspoints
   \advance\epsfrsize by-\epsflly\pspoints
   \epsftsize=\epsfurx\pspoints
   \advance\epsftsize by-\epsfllx\pspoints
   \epsfxsize\epsfsize\epsftsize\epsfrsize
   \ifnum\epsfxsize=0 \ifnum\epsfysize=0
      \epsfxsize=\epsftsize \epsfysize=\epsfrsize
     \else\epsftmp=\epsftsize \divide\epsftmp\epsfrsize
       \epsfxsize=\epsfysize \multiply\epsfxsize\epsftmp
       \multiply\epsftmp\epsfrsize \advance\epsftsize-\epsftmp
       \epsftmp=\epsfysize
       \loop \advance\epsftsize\epsftsize \divide\epsftmp 2
       \ifnum\epsftmp>0
          \ifnum\epsftsize<\epsfrsize\else
             \advance\epsftsize-\epsfrsize \advance\epsfxsize\epsftmp \fi
       \repeat
     \fi
   \else\epsftmp=\epsfrsize \divide\epsftmp\epsftsize
     \epsfysize=\epsfxsize \multiply\epsfysize\epsftmp   
     \multiply\epsftmp\epsftsize \advance\epsfrsize-\epsftmp
     \epsftmp=\epsfxsize
     \loop \advance\epsfrsize\epsfrsize \divide\epsftmp 2
     \ifnum\epsftmp>0
        \ifnum\epsfrsize<\epsftsize\else
           \advance\epsfrsize-\epsftsize \advance\epsfysize\epsftmp \fi
     \repeat     
   \fi
   \ifepsfverbose\message{#1: width=\the\epsfxsize, height=\the\epsfysize}\fi
   \epsftmp=10\epsfxsize \divide\epsftmp\pspoints
   \vbox to\epsfysize{\vfil\hbox to\epsfxsize{%
      \includegraphics{#1}%
      \hfil}}%
\epsfxsize=0pt\epsfysize=0pt}%

{\catcode`\%=12 \global\let\epsfpercent=%\global\def\epsfbblit{%BoundingBox}}%
\long\def\epsfaux#1#2:#3\\{\ifx#1\epsfpercent
   \def\testit{#2}\ifx\testit\epsfbblit
      \epsfgrab #3 . . . \\%
      \epsffileokfalse
      \global\epsfbbfoundtrue
   \fi\else\ifx#1\par\else\epsffileokfalse\fi\fi}%
\def\epsfgrab #1 #2 #3 #4 #5\\{%
   \global\def\epsfllx{#1}\ifx\epsfllx\empty
      \epsfgrab #2 #3 #4 #5 .\\\else
   \global\def\epsflly{#2}%
   \global\def\epsfurx{#3}\global\def\epsfury{#4}\fi}%
\def\epsfsize#1#2{\epsfxsize}
\let\epsffile=\epsfbox

\newlength{\dinwidth}
\newlength{\dinmargin}
\setlength{\dinwidth}{21.0cm}
\textheight24.2cm \textwidth17.0cm
\setlength{\dinmargin}{\dinwidth}
\addtolength{\dinmargin}{-\textwidth}
\setlength{\dinmargin}{0.5\dinmargin}
\oddsidemargin -1.0in
\addtolength{\oddsidemargin}{\dinmargin}
 
\setlength{\evensidemargin}{\oddsidemargin}
\setlength{\marginparwidth}{0.9\dinmargin}
\marginparsep 8pt \marginparpush 5pt
\topmargin -42pt
\headheight 12pt
\headsep 30pt %\footheight 12pt \footskip 24pt
\parskip 3mm plus 2mm minus 2mm
\parindent 5mm
\renewcommand{\arraystretch}{1.3}
\renewcommand{\thefootnote}{\arabic{footnote}}

\newcommand{\bfig}{\begin{figure}}
\newcommand{\efig}{\end{figure}}
\newcommand{\bcen}{\begin{center}}
\newcommand{\ecen}{\end{center}}
\newcommand{\beq}{\begin{equation}}
\newcommand{\eeq}{\end{equation}}
\newcommand{\btabu}{\begin{tabular}}
\newcommand{\etabu}{\end{tabular}}
\newcommand{\btabl}{\begin{table}}
\newcommand{\etabl}{\end{table}}

\newcommand{\GeV}{\mbox{GeV}}

\def\lsim{\mathrel{\rlap{\lower4pt\hbox{\hskip1pt$\sim$}}
    \raise1pt\hbox{$<$}}}         %less than or approx. symbol
\def\gsim{\mathrel{\rlap{\lower4pt\hbox{\hskip1pt$\sim$}}
    \raise1pt\hbox{$>$}}}         %greater than or approx. symbol

\begin{document}
\title{\bf 
Measurement of the {\boldmath $t$} Distribution in Diffractive
Photoproduction at HERA}
\vspace{5cm}
\author{\Large The ZEUS Collaboration} 
\date{}

\maketitle
\thispagestyle{empty}
\vspace{3cm}
\begin{abstract}
Photon diffractive dissociation, $\gamma p \rightarrow Xp$, has been
studied at HERA with the ZEUS detector using $ep$ interactions where the 
virtuality $Q^2$ of the exchanged photon is smaller than 0.02~GeV$^2$.
The squared four-momentum $t$ exchanged at the proton vertex was
determined in the range $0.073<|t|<0.40$~GeV$^2$ by measuring the 
scattered proton in the ZEUS Leading Proton Spectrometer. 
In the photon-proton centre-of-mass energy interval
$176<W<225$~GeV and for masses of the dissociated photon system $4<M_X<32$~GeV,
the $t$ distribution has an exponential shape, 
$dN/d|t| \propto \exp{(-b|t|)}$, with a slope 
parameter $b=6.8 \pm 0.9$~(stat.)~$ ^{+1.2}_{-1.1}$~(syst.)~GeV$^{-2}$.
\end{abstract}

\vspace{-16.5cm}
\begin{flushleft}
\tt DESY 97-238 \\
November 1997 \\
\end{flushleft}

\thispagestyle{empty}
\newpage

%===================================================================                               
%                                                                                                  
%  MEMBER NAME  AUTH57 (ZEUS)     M  TEX                                                           
%                                                                                                  
%  JH.: transformed to a format, which is suited as input for                                      
%       CONVERT, which automatically creates author-indices                                        
%                                                                                                  
%  Don't remove lines starting with a percent sign %,                                              
%  CONVERT may need them urgently !                                                                
%                                                                                                  
%=====================================================================                             
%\documentstyle[twoside]{report}                                                                    
\topmargin-1.cm                                                                                    
\evensidemargin-0.3cm                                                                              
\oddsidemargin-0.3cm                                                                               
\textwidth 16.cm                                                                                   
\textheight 680pt                                                                                  
\parindent0.cm                                                                                     
\parskip0.3cm plus0.05cm minus0.05cm                                                               
\def\3{\ss}                                                                                        
\newcommand{\address}{ }                                                                           
\renewcommand{\author}{ }                                                                          
\pagenumbering{Roman}                                                                              
                                    % this "%"s are for cosmetics only                             
%\begin{document}                                                                                   
                                                   %                                               
\begin{center}                                                                                     
{                      \Large  The ZEUS Collaboration              }                               
\end{center}                                                                                       
  J.~Breitweg,                                                                                     
  M.~Derrick,                                                                                      
  D.~Krakauer,                                                                                     
  S.~Magill,                                                                                       
  D.~Mikunas,                                                                                      
  B.~Musgrave,                                                                                     
  J.~Repond,                                                                                       
  R.~Stanek,                                                                                       
  R.L.~Talaga,                                                                                     
  R.~Yoshida,                                                                                      
  H.~Zhang  \\                                                                                     
 {\it Argonne National Laboratory, Argonne, IL, USA}~$^{p}$
% ...
\newpage                                                                                           
$^{\    1}$ also at IROE Florence, Italy \\                                                        
$^{\    2}$ now at Univ. of Salerno and INFN Napoli, Italy \\                                      
% ...
$^{  25}$ now at ATLAS Collaboration, Univ. of Munich \\                                           
% ...
$^{  34}$ partially supported by the Foundation for German-Russian Collaboration                   
DFG-RFBR \\ \hspace*{3.5mm} (grant no. 436 RUS 113/248/3 and no. 436 RUS 113/248/2)\\              
% ...
                                                           %                                       
                                                           %                                       
% \par         % if index listing & table fit to 1 page, put gap here                              
\newpage   % alternatively: go to newpage, if page is too small                                    
                                                           %                                       
% \institute_references_start    % do not touch or move this line !                                
                                                           %                                       
\begin{tabular}[h]{rp{14cm}}                                                                       
$^{a}$ &  supported by the Natural Sciences and Engineering Research                               
          Council of Canada (NSERC)  \\                                                            
$^{b}$ &  supported by the FCAR of Qu\'ebec, Canada  \\                                            
$^{c}$ &  supported by the German Federal Ministry for Education and                               
          Science, Research and Technology (BMBF), under contract                                  
          numbers 057BN19P, 057FR19P, 057HH19P, 057HH29P, 057SI75I \\                              
$^{d}$ &  supported by the MINERVA Gesellschaft f\"ur Forschung GmbH,                              
          the German Israeli Foundation, and the U.S.-Israel Binational                            
          Science Foundation \\                                                                    
$^{e}$ &  supported by the German Israeli Foundation, and                                          
          by the Israel Science Foundation                                                         
  \\                                                                                               
$^{f}$ &  supported by the Italian National Institute for Nuclear Physics                          
          (INFN) \\                                                                                
$^{g}$ &  supported by the Japanese Ministry of Education, Science and                             
          Culture (the Monbusho) and its grants for Scientific Research \\                         
$^{h}$ &  supported by the Korean Ministry of Education and Korea Science                          
          and Engineering Foundation  \\                                                           
$^{i}$ &  supported by the Netherlands Foundation for Research on                                  
          Matter (FOM) \\                                                                          
$^{j}$ &  supported by the Polish State Committee for Scientific                                   
          Research, grant No.~115/E-343/SPUB/P03/002/97, 2P03B10512,                               
          2P03B10612, 2P03B14212, 2P03B10412 \\                                                    
$^{k}$ &  supported by the Polish State Committee for Scientific                                   
          Research (grant No. 2P03B08308) and Foundation for                                       
          Polish-German Collaboration  \\                                                          
$^{l}$ &  partially supported by the German Federal Ministry for                                   
          Education and Science, Research and Technology (BMBF)  \\                                
$^{m}$ &  supported by the Fund for Fundamental Research of Russian Ministry                       
          for Science and Edu\-cation and by the German Federal Ministry for                       
          Education and Science, Research and Technology (BMBF) \\                                 
$^{n}$ &  supported by the Spanish Ministry of Education                                           
          and Science through funds provided by CICYT \\                                           
$^{o}$ &  supported by the Particle Physics and                                                    
          Astronomy Research Council \\                                                            
$^{p}$ &  supported by the US Department of Energy \\                                              
$^{q}$ &  supported by the US National Science Foundation \\                                       
\end{tabular}                                                                                      

\newpage

\setlength{\dinwidth}{21.0cm}
\textheight24.2cm \textwidth17.0cm
\setlength{\dinmargin}{\dinwidth}
\addtolength{\dinmargin}{-\textwidth}
\setlength{\dinmargin}{0.5\dinmargin}
\oddsidemargin -1.0in
\addtolength{\oddsidemargin}{\dinmargin}
 
\setlength{\evensidemargin}{\oddsidemargin}
\setlength{\marginparwidth}{0.9\dinmargin}
\marginparsep 8pt \marginparpush 5pt
\topmargin -42pt
\headheight 12pt
\headsep 30pt %\footheight 12pt \footskip 24pt
\parskip 3mm plus 2mm minus 2mm
\parindent 5mm
\renewcommand{\arraystretch}{1.3}
\renewcommand{\thefootnote}{\arabic{footnote}}
\pagenumbering{arabic} 
\setcounter{page}{1}

\section{Introduction}

The reaction $\gamma p \rightarrow Xp$, in which the photon
diffractively dissociates into an hadronic state $X$ with mass $M_X$, 
has been investigated with real photons at a photon-proton 
centre-of-mass energy $W$ of about 14~GeV~\cite{chapin}. Recently it 
has also been studied at HERA using the process 
$ep \rightarrow eXp$ for photon virtualities $Q^2<0.02$~GeV$^2$ and 
$W \approx 200$~GeV~\cite{h1_mx,zeus_mx}. 
The comparison of the fixed target data and the HERA data indicates that the 
dissociation of real photons has
similar characteristics to the dissociation of hadrons, as expected in the 
framework of Vector Meson Dominance (VMD)~\cite{sakurai,bauer}. In this 
model, the photon is assumed to fluctuate into a virtual vector meson 
prior to the interaction with the proton. The interaction can be 
described by Regge phenomenology~\cite{collins} and, at high energy, 
is dominated by the 
exchange of an object with the quantum numbers of the vacuum,
referred to as the pomeron. An exponential fall of the differential cross 
section $d\sigma/d|t|\propto \exp{(-b|t|)}$, at small values 
of $|t|$, is a 
typical feature of diffraction; here $t$ is the square of the 
four-momentum transfer at the proton vertex. Regge theory also predicts 
that the diffractive peak shrinks as $W$ increases according to
$b = b_0+ 2 \alpha^{\prime}\ln{(W^2/M_X^2)}$, 
where $b_0$ and $\alpha^{\prime}$ are 
constants~\cite{collins,field_fox}.

The studies of diffractive real photon dissociation at HERA have 
so far focussed on the shape of the $M_X$ spectrum~\cite{h1_mx,zeus_mx}.
The $t$ distribution for the reaction $\gamma p \rightarrow X p$ has been
measured only by the fixed target experiment~\cite{chapin}, which 
found that in the range $1.4<M_X<3$~GeV and for $0.02<|t|<0.1$~GeV$^2$
the $t$ dependence is exponential with a $t$-slope $b \approx 5$~GeV$^{-2}$.
At HERA, measurements of the $t$ distribution 
have been performed for the diffractive dissociation of virtual photons
in the range $5<Q^2<20$~GeV$^2$~\cite{lps_f2d3}, and for elastic vector meson 
production, $\gamma p \rightarrow Vp$, both with real and with virtual 
%photons~\cite{rho93}-\cite{psi_zeus}. 
photons~[9-18]. 
In all cases the distribution has an approximately exponential shape. 
The $t$-slope is 
$b= 7.2 \pm 1.1$~(stat.)~$^{+0.7}_{-0.9}$~(syst.)~GeV$^{-2}$
for the diffractive dissociation 
of virtual photons at $\langle Q^2 \rangle=8$~GeV$^2$. For elastic $\rho^0$ 
production $b$ depends only weakly on $W$ but varies from 
%approximately $10$~GeV$^{-2}$ for $Q^2 \approx 0$~\cite{rho93}-\cite{lps_rho} 
approximately $10$~GeV$^{-2}$ for $Q^2 \approx 0$~[9-11]
to approximately 
5-7~GeV$^{-2}$ for 
$\langle Q^2 \rangle=10$~GeV$^2$~\cite{rhodis_zeus,rhodis_h1}. 
It is therefore interesting to extend these measurements to 
diffractive real photon dissociation.

In this paper we report the first determination at HERA of the $t$ 
distribution for the process $\gamma p \rightarrow Xp$, where $\gamma$ is a 
photon with $Q^2<0.02$~GeV$^2$.
The present measurement is based on a sample of photoproduction events
collected using the reaction $ep \rightarrow eXp$ at 
$W \approx 200$~GeV~\cite{zeus_mx}. 
The sample was defined by the 
requirement that the scattered positron be measured in a calorimeter close 
to the outgoing positron beam line and a final state proton 
carrying at least 97\% of the incoming proton momentum be detected in 
the ZEUS Leading Proton Spectrometer (LPS)~\cite{lps_rho}. 
The LPS was also used to measure the transverse momentum of the proton, 
from which $t$ was calculated. This is a technique similar to that used to 
measure the $t$ distribution in the photoproduction of $\rho^0$ mesons, 
$\gamma p \rightarrow \rho^0p$~\cite{lps_rho}, and for the diffractive
dissociation of virtual photons, 
$\gamma^* p \rightarrow Xp$~\cite{lps_f2d3}.

\section{Experimental set-up}
\label{setup}

\subsection{HERA}

The data presented here were collected in 1994 at 
HERA which operated with 820~GeV protons and 27.5~GeV positrons. The 
proton and positron beams each contained 153 colliding bunches, 
together with 17 additional unpaired proton and 15 unpaired positron 
bunches. These 
additional bunches were used for background studies. 
The integrated luminosity for the 
present study, which required the LPS to be in operation,
is 0.9~pb$^{-1}$.

\subsection{The ZEUS detector}

A detailed description of the ZEUS detector can be found 
elsewhere~\cite{detector_a,detector_b}. A brief outline of 
the components which are most relevant for this analysis is given below.
Throughout this paper the standard ZEUS coordinate system is used,
which has the origin at the nominal interaction point, the
$Z$ axis pointing in the proton beam direction, hereafter referred to 
as ``forward'', the $X$ axis pointing horizontally towards the centre of 
HERA and the $Y$ axis pointing upwards. The polar angle
$\theta$ is defined with respect to the  $Z$ direction.

Charged particles are measured by the inner tracking detectors which operate in a
magnetic field of 1.43 T provided by a thin superconducting solenoid.
Immediately surrounding the beam-pipe is the vertex detector (VXD), 
a drift chamber which consists of 120 radial cells, each with 12 
sense wires~\cite{vxd}. 
It is surrounded in turn by the central tracking detector (CTD), which consists 
of 72 cylindrical drift chamber layers,  organized into 9 
superlayers covering the polar angle region 
$15^\circ < \theta < 164^\circ$~\cite{ctd}.

For the energy measurement the high resolution
depleted-uranium scintillator calorimeter (CAL) is used~\cite{CAL}.
It is divided
into three parts, forward (FCAL) covering the
pseudorapidity\footnote{The pseudorapidity \mbox{$\eta$} is defined as
$ \eta = - \ln{(\tan{(\theta/2)})}$.} 
region \mbox{$4.3>\eta>1.1$}, barrel (BCAL) covering the central region
\mbox{$1.1>\eta>-0.75$} and rear (RCAL) covering the backward region
\mbox{$-0.75>\eta>-3.8$}.  Holes of $20 \times 20$~cm$^{2}$ in
the centre of FCAL and RCAL accommodate the HERA beam-pipe.
Each of the calorimeter parts is subdivided into towers, which in
turn are segmented longitudinally into electromagnetic (EMC) and
hadronic (HAC) sections.  These sections are further subdivided
into cells, which are read out by two photomultiplier tubes.
Under test beam conditions, the energy resolution of the calorimeter
was measured to be \mbox{$\sigma_{E}/E = 0.18/\sqrt{E (\GeV)}$} for electrons 
and \mbox{$\sigma_{E}/E = 0.35/\sqrt{E (\GeV)}$} for hadrons.
The calorimeter noise, dominated by the uranium radioactivity, is in
the range 15-19~MeV for an EMC cell and 24-30~MeV for a HAC
cell. 

The luminosity is determined from the rate of the Bethe-Heitler 
process, $ep \rightarrow e \gamma p$, where the photon is measured with a 
lead-scintillator sandwich calorimeter (LUMI-$\gamma$) located 
at $Z= - 107$~m in the HERA tunnel downstream of the 
interaction point in the direction of the outgoing positrons~\cite{lumi}.
A similar calorimeter (LUMI-$e$) at $Z=-35$~m detects
positrons scattered at very small angles.
In this analysis, the LUMI-$e$ was used to tag photoproduction
events with positrons scattered at angles up to about 5~mrad and to measure
the scattered positron energy, $E'_{e}$. The LUMI-$e$ covers the range
$7<E'_{e}<21$~GeV.
The energy resolution of both calorimeters is 
\mbox{$\sigma_E/E =0.18/\sqrt{E (\GeV)}$}.

The Leading Proton Spectrometer (LPS)~\cite{lps_rho} detects
charged particles scattered at small angles and carrying a substantial 
fraction, $x_L$, of the incoming proton momentum; these particles remain 
in the beam-pipe and their trajectory is measured by a system of 
silicon micro-strip detectors very close (typically a few mm) 
to the proton beam. The detectors are grouped in six stations, 
S1 to S6, placed along the beam line in the direction of the outgoing 
protons, at $23.8$~m, 40.3~m, 44.5~m, 
63.0~m, 81.2~m and 90.0~m from the interaction point.
The track deflections induced by the magnets in the 
proton beam line allow a momentum analysis of the scattered proton.
For the present measurements, only the stations S4, S5 and S6 were used.
With this configuration, for $x_L$ close to unity, resolutions of 0.4\% 
on the longitudinal 
momentum and 5~MeV on the transverse momentum have been achieved. 
The effective transverse momentum
resolution is, however, dominated by the intrinsic transverse 
momentum spread of the proton beam at the interaction point 
which is $\approx 40$~MeV in the horizontal plane and $\approx 90$~MeV in the vertical 
plane.
For $x_L$ close to unity, the LPS covers the range
$0.25 \lsim p_T \lsim 0.65$~GeV, where 
$p_T$ is the transverse momentum of the proton with respect to
the incoming beam direction. As discussed previously~\cite{lps_rho}, the incoming
beam direction and the beam position with respect to each station
are determined using the reaction $ep \rightarrow e \rho^0 p$ at $Q^2 
\approx 0$.
Protons with $p_T<0.2$~GeV and $x_L\approx 1$ are too close to the beam to be 
measured. For the events considered here the 
geometric acceptance of the LPS is approximately 6\%.

\section{Data selection and background subtraction}

\subsection{Trigger}

ZEUS uses a three-level trigger system~\cite{detector_a,detector_b}. 
At the first-level a coincidence between signals in the LUMI-$e$ 
and in the RCAL was required. An energy deposit greater than 5~GeV
was required in the LUMI-$e$. In the RCAL the deposit had to be larger 
than 464~MeV (excluding the towers immediately adjacent to the beam-pipe) 
or 1250~MeV (including those towers).
The angular acceptance of the LUMI-$e$ limits the $Q^2$ range 
to the region $Q^2<0.02$~GeV$^2$. The small RCAL threshold essentially 
selects all photoproduction events. The second and third trigger levels
were mainly used to reject beam related background.
Parts of the data stream were 
prescaled~\cite{zeus_mx,michal} in order to reduce the high event rate 
resulting from the large photoproduction cross section.

\subsection{Reconstruction of variables}

The photon-proton centre-of-mass energy squared, $W^2=(q+p)^2$, where $q$
and $p$ are the virtual photon and the proton four-momenta, respectively,
was determined by $W^2\approx ys$, with 
$y \approx E_{\gamma}/E_e = (E_e-E_e^{\prime})/E_e$ and
$s$ the squared centre-of-mass energy of the positron-proton system;
here $E_{\gamma}$ is the energy of the exchanged photon and $E_e$ 
denotes the energy of the incoming positron.
The $W$ resolution is 7~GeV at $W=176$~GeV and improves to 4.5~GeV at 
$W=225$~GeV.

The mass $M_X$ of the dissociated photon system
was reconstructed~\cite{zeus_mx} by combining the information from the LUMI-$e$ 
and the CAL:
\begin{equation}
M_{X}  = \sqrt{E^{2}-P^{2}} \approx  \sqrt{(E-P_{Z}) \cdot (E+P_{Z})} =
\sqrt{2 E_{\gamma} \cdot (E+P_{Z}), }
\end{equation}
\noindent
where $E$ is the energy of the hadronic system observed in the CAL; 
the total momentum of the hadronic system, $P$,
approximately equals the longitudinal component, $P_Z$, as the
transverse component generally is small in the case of photoproduction events.
The following formula was used for the mass reconstruction:
\begin{equation}
M_{X\:rec} = a_1 \cdot
\sqrt{ 2E_{\gamma} \cdot
(\sum_{cond}E_i + \sum_{cond}E_i \cos\theta_i )}+a_2.
\label{e:mx_rec}
\end{equation}
\noindent
The quantities $E_i$ and $\theta_i$ denote the energy and the polar angle of
CAL condensates, defined as groups of adjacent cells with total energy
of at least 100~MeV, if all the cells belong to the EMC, or 200~MeV 
otherwise. These cuts reduce the effect of noise on the mass 
reconstruction. They were applied in addition to a noise suppression 
algorithm which discarded all EMC (HAC) cells with energy below 
60~MeV (110~MeV); for isolated cells the thresholds were increased to
80~MeV (120~MeV). 
The coefficients $a_1$ and $a_2$ correct for the effects of energy loss in
the inactive material in front of the CAL and of energy deposits below the 
threshold.
Their values, $a_1=1.14$ and $a_2=1.2$~GeV, were taken from~\cite{zeus_mx}.
The masses in the range $4<M_X<40$~GeV are reconstructed with a 
resolution 
\mbox{$\sigma_{M_X}/M_X \approx 0.8/\sqrt{M_X (\GeV)}$} and an offset
smaller than 0.5~GeV~\cite{zeus_mx}.

The variable $t=(p-p^{\prime})^2$, where $p$ and
$p^{\prime}$ are the incoming and the scattered proton four-momenta,
respectively, can be evaluated as 
$t\approx -(p_T^2/x_L) [1+(M_p^2/p_T^2)(1-x_L)^2]$.
Both $p_T$ and $x_L$
were measured with the LPS.
For the data considered here, which have $x_L$ close to 
unity, the approximation $t \approx -p_T^2/x_L$ was used.
Since, as mentioned earlier, the incoming proton beam has an intrinsic transverse momentum spread of 
$\sigma_{p_X} \approx 40$~MeV and 
$\sigma_{p_Y} \approx 90$~MeV, which is much 
larger than the LPS resolution in transverse momentum, the measured 
value of $t$ is given by the convolution of the true $t$ 
distribution and the effect of the beam spread. Because of this 
we make a distinction between the true value of
$t$ and the measured value, $t_{apparent}=-p_T^2/x_L$.

\subsection{Offline selection}

To select the final sample, the following conditions were imposed 
on the reconstructed data:

\begin{itemize}
\item A scattered positron in the LUMI-$e$ 
with energy in the range $12<E_e^{\prime}<18$~GeV,
corresponding to $176<W<225$~GeV.

\item An interaction vertex reconstructed by the tracking detectors.

\item Mass of the dissociated photon system in the range $4<M_{X \:rec}<32$~GeV.
The lower limit eliminates the region dominated by resonance production; it 
also reflects the lower limit of $M_X=1.7$~GeV for which Monte Carlo events
were generated (cf. section~\ref{montecarlo}).
The upper limit is a consequence of the limit on $x_L$ 
(see below), since $M_X^2 \approx W^2(1-x_L)$.

\end{itemize}

\noindent

In addition, the detection of a high momentum proton in the 
LPS was required~\cite{lps_rho}:

\begin{itemize}
\item One track in the LPS with $x_L>0.97$ was required.
This is used to select diffractive events in which the proton remains intact.

\item Protons with reconstructed trajectories closer than 0.5~mm
to the wall of the beam-pipe, at any point between the vertex and the last
station hit, were rejected. 
This eliminates any
sensitivity of the acceptance to possible misalignments of the HERA beam-pipe 
elements. In addition badly reconstructed tracks are removed.

\item The $p_T$ range was restricted to the interval $0.27<p_T<0.63$~GeV,
thereby removing regions where the acceptance of the LPS is very small or 
changes rapidly~\cite{lps_rho}.
               
\end{itemize}

\noindent
After these selections, 641 events remained.

\subsection{Background}

The background contamination in the sample was mainly due to two
sources.
\begin{enumerate}

\item 
Some activity in the RCAL can accidentally overlap with the scattered 
positron of a bremsstrahlung event 
($ep \rightarrow e \gamma p$) in the LUMI-$e$~\cite{zeus_mx}.
A large fraction of this background can be identified since
the bremsstrahlung photon is accepted by the LUMI-$\gamma$.
For bremsstrahlung, one has 
$E_e^{\prime}+E_{\gamma}=E_e$, where $E_{\gamma}$ is the energy of 
the radiated photon. For such events the energy deposits in the 
LUMI-$e$ and the LUMI-$\gamma$ calorimeters thus sum up to the positron beam 
energy. These events were removed.

The unidentified events were statistically subtracted by
including the identified background events with negative 
weights in all the distributions, thereby compensating for the unidentified 
background events~\cite{ZEUS_sigtot,Burow}.
This subtraction was less than 3\%.

\item 
A proton beam-halo track in the LPS can accidentally overlap with
an event satisfying the trigger and the selection cuts
applied to the variables measured with 
the central ZEUS detector (beam-halo event).
The term beam-halo track refers to a proton with
energy close to that of the beam originating from
an interaction of a beam proton with the residual gas in the
pipe or with the beam collimator jaws.
Obviously, a beam-halo track is uncorrelated with the activity
in the central ZEUS detector. For such a beam-halo event, energy and momentum 
conservation are not necessarily satisfied; in particular the quantity
$(E+P_Z+2P_Z^{LPS})$, where $P_Z^{LPS}$ is the $Z$ component of the
proton momentum measured in the LPS, may exceed the kinematic limit of 
1640~GeV. The condition $(E+P_Z+2P_Z^{LPS})> 1655$~GeV 
(thereby including the effects of 
resolution) identifies such events, which thus were rejected. 

In order to evaluate the residual contamination after all cuts, 
the distribution of $2P_Z^{LPS}$ for 
identified beam-halo events was randomly mixed with the 
$(E+P_Z)$ distribution for all events, so as to create a distribution of 
$(E+P_Z+2P_Z^{LPS})$ for halo events. 
The observed $(E+P_Z+2P_Z^{LPS})$ distribution was then fitted 
as the sum of the diffractive Monte Carlo contribution 
(cf.~section~\ref{montecarlo}) and the beam-halo contribution just discussed.
The relative normalisation of the two terms was left as a free 
parameter. 
For events with $(E+P_Z+2P_Z^{LPS})< 1655$~GeV,
the fraction of beam-halo events
thus was estimated to be $(6.3\pm1.2)\%$.
Here again the identified background events were included with negative
weights in all the distributions in order to compensate 
for the unidentified beam-halo events.

\end{enumerate}

The number of events remaining after the background subtraction, i.e.
after removing the identified bremsstrahlung and beam-halo events
and after including the effect of the negative weights, was 515. 
The contribution from non-single diffractive dissociation processes, 
e.g. double diffractive dissociation, 
is expected to be of the order of a few per
cent~\cite{lps_f2d3} and was not subtracted.

\begin{figure}
\vspace{-2.4cm}
\begin{center}
\leavevmode
\hbox{%
\epsfxsize = 15cm
\epsffile{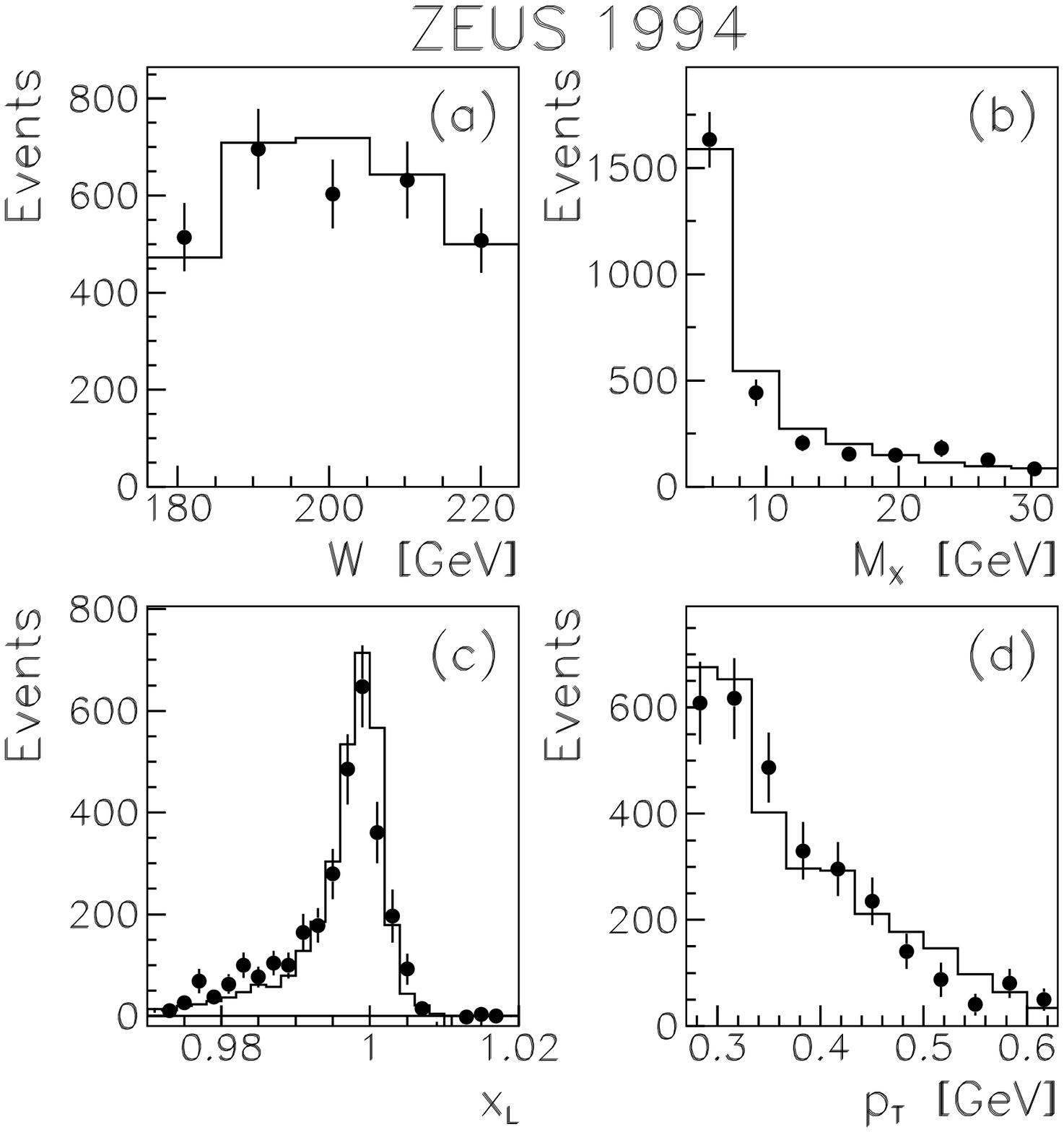}}
\end{center}
\vspace{1cm}
\caption{Distributions as a function of 
(a) $W$, (b) $M_X$, (c) $x_L$ and (d) $p_T$ for
the data (points) and the Monte Carlo events
(histogram). 
The error bars indicate the statistical uncertainty of the data.
The number of data events is corrected for the trigger 
prescale factors and the background.
}
\label{data_mc}
\end{figure}

Figure~\ref{data_mc} shows the observed $W$, $M_X$, $x_L$ and 
$p_T$ distributions for the selected events after background 
subtraction and including the correction for the effects of the trigger 
prescale factors. 

\section{Monte Carlo simulation and acceptance determination}
\label{montecarlo}

The reaction $ep \rightarrow eXp$ was simulated using a Monte Carlo 
generator~\cite{ada} based on a model calculation by Nikolaev and 
Zakharov~\cite{nik_zak}. The generated $M_X$ distribution was reweighted 
with the sum of a pomeron-pomeron-pomeron~\cite{field_fox} %($\pom\pom\pom$) 
contribution
($d\sigma/dM_X^2 \propto 1/M_X^2$) and a 
pomeron-pomeron-reggeon~\cite{field_fox} 
contribution
($d\sigma/dM_X^2 \propto 1/M_X^3$), so as to obtain a satisfactory 
agreement between data and Monte Carlo.
As discussed in sect.~\ref{t_dist}, however, the present results on 
the $t$ distribution are
largely independent of the details of the $M_X$ spectrum simulation
in the mass range considered in the analysis.

All generated events were passed through the standard ZEUS detector
simulation, based on the GEANT program~\cite{geant}, and through
the trigger simulation package.
The simulation also includes the geometry of the beam-pipe apertures,
the HERA magnets and their fields.
The spread of the interaction vertex position was also simulated and so were
the proton beam angle with respect to the nominal direction and its dispersion
at the interaction point.
The simulated events were then passed through the 
same reconstruction and analysis programs as the data. 
In Fig.~\ref{data_mc}
the distributions for the reconstructed Monte Carlo 
events as a function of $W$, $M_X$, $x_L$ and $p_T$ 
are compared with those of the data. The distributions of the simulated 
events were normalised to the observed number of events, 
corrected for the effects of the prescale and of the background 
subtraction. The agreement between the data and the Monte Carlo
distributions is satisfactory.

The acceptance was computed as the ratio of the number of reconstructed 
Monte Carlo events in a bin of a given variable and the number of generated events in that bin.
The acceptance thus includes the effects of the geometric acceptance of the apparatus, 
its efficiency and resolution, as well as the trigger and reconstruction 
efficiencies.

\section{Results and discussion}
\label{t_dist}

\begin{figure}
\vspace{-2.4cm}
\begin{center}
\leavevmode
\hbox{%
\epsfxsize = 15cm
\epsffile{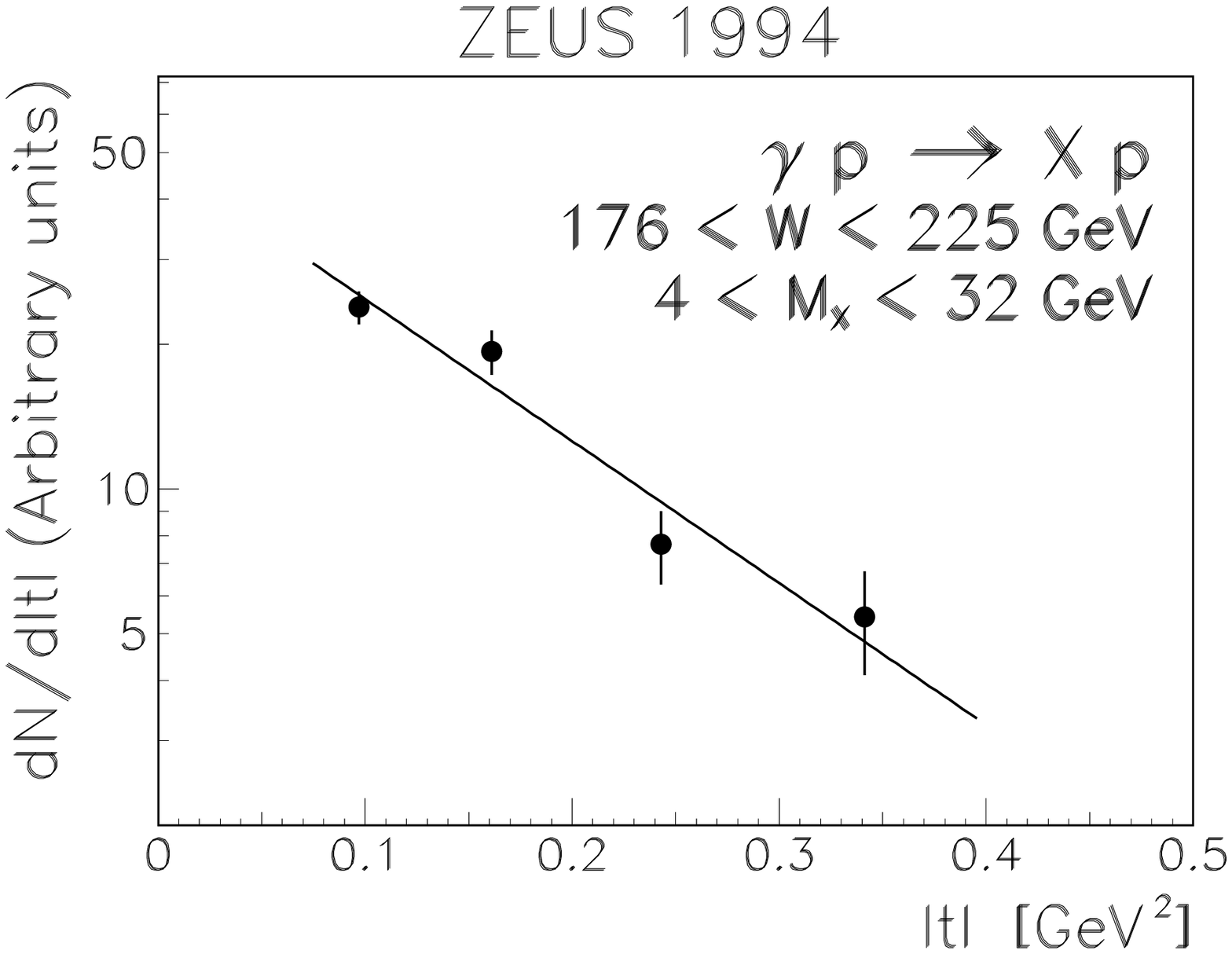}}
\end{center}
\vspace{1cm}
\caption{
Differential distribution $dN/d|t|$
for photon diffractive dissociation, $\gamma p \rightarrow X p$,
in the kinematic region $176< W< 225$~GeV 
and $4<M_X<32$~GeV.
The vertical bars indicate the size of the statistical uncertainties.
The line is the result of the fit described in the text. The scale on the
vertical axis is arbitrary.
}
\label{dndt}
\end{figure}

The acceptance corrected $t$ distribution, $dN/d|t|$, is shown in Fig.~\ref{dndt}.
It was obtained by correcting the measured $t_{apparent}$ distribution
bin by bin with the acceptance determined from the Monte Carlo simulation described 
above.

The data were fitted with the function:

\begin{eqnarray}
\frac{ dN}{d|t|}   = A     \cdot e^{-b |t|},
\label{single}
\end{eqnarray}
\noindent
where $A$ is a constant. The resulting value of the $t$-slope is
\begin{eqnarray}
b = 6.8\pm 0.9~(\mbox{stat.})^{+1.2}_{-1.1}~(\mbox{syst.})~\mbox{GeV}^{-2},\nonumber
\label{result}
\end{eqnarray}
\noindent
in the kinematic region $4<M_X<32$~GeV, $0.073<|t|<0.40$~GeV$^2$
and $176<W<225$~GeV. In this region, the average value of $W$ is 200~GeV and 
the average value of $M_X$ is 11~GeV. The result of the fit is indicated by
the continuous line on Fig.~\ref{dndt}.

The analysis was repeated for the two 
$M_X$ ranges $4<M_X<8$~GeV and $8<M_X<32$~GeV. The results,
$b=7.0 \pm 1.3$~GeV$^{-2}$ and 
$b=6.5 \pm 1.3$~GeV$^{-2}$, indicate no
variation with $M_X$ within the present sensitivity.

The quoted systematic uncertainty on $b$ ($\Delta b$)
was obtained by modifying the requirements
and the analysis procedures as listed below:

\begin{enumerate}

\item Sensitivity to the selection of the proton track~(cf.~\cite{lps_rho}):

\begin{itemize}

\item The sensitivity to the proton beam 
tilt with respect to the nominal 
was evaluated by systematically shifting $p_T$ by 10~MeV. 

\item The track selection requirements were tightened.

\item Events with $p_X^{LPS}>0$ and with $p_X^{LPS}<0$ were analysed 
separately, as a check of possible relative rotations of the LPS 
detector stations.

\item The data were divided into a ``low acceptance" and
a ``large acceptance" sample depending on the position of the
LPS stations. The latter varied slightly from run to run.

\end{itemize}

The last three contributions dominate; by summing all four
in quadrature, $\Delta b = \pm 1.0$~GeV$^{-2}$ was obtained. 

\item Sensitivity to the other selection cuts and acceptance corrections:

\begin{itemize}

\item The $W$ range was restricted to $195<W<215$~GeV, leading to 
$b=6.7\pm 1.4$~GeV$^{-2}$.

\item The $M_X$ distribution in the Monte Carlo was varied between
 $d\sigma/dM_X^2 \propto (1/M_X)^{1.5}$ and $d\sigma/dM_X^2 \propto 
 (1/M_X)^{3}$. The corresponding variation 
 of $b$ was at most $\pm 0.2$~GeV$^{-2}$.

\item The vertex requirement was removed or restricted to 
$|Z_{vertex}|<50$~cm. The effect on $b$ was at most $\pm 0.4$~GeV$^{-2}$.

\end{itemize}

Summing these contributions to $\Delta b$ in quadrature
yields $\Delta b=\pm 0.5$~GeV$^{-2}$. 

\item Background corrections: the size of the beam-halo background was
varied by two standard deviations, yielding negligible effects. The 
bremsstrahlung background correction was removed altogether, 
causing changes in the result smaller than 0.1~GeV$^{-2}$. 

\item Evaluation of the $t$-slope:
\begin{itemize}

 \item The $t$-slope was determined with an alternative method discussed in 
 detail in refs.~\cite{lps_rho,thesis_roberto}. One can 
 express the $t_{apparent}$ distribution as a
 convolution of equation~(\ref{single}) and a two-dimensional Gaussian
 distribution representing the beam transverse momentum distribution. 
  The slope parameter $b$ can then be determined by fitting the convolution of 
 eq.~(\ref{single}) and the two-dimensional Gaussian to the acceptance 
corrected 
 $t_{apparent}$ distribution. This method has the advantage that the 
 data can be binned in $t_{apparent}$ for which 
 the resolution is much better than for $t$, as discussed earlier.
 The value of $b$ thus obtained was $b=7.3\pm 0.9$~GeV$^{-2}$.

 \item The $t$-slope was also obtained with a third method: the 
 $t_{apparent}$ distributions in the data and in the Monte Carlo were 
 compared and the Monte Carlo $t$ distribution at the generator level was reweighted until the
 $\chi^2$ of the comparison between data and Monte Carlo reached a minimum. 
 The result thus found differed from the nominal one by less than 
0.1~GeV$^{-2}$.

 \item The sensitivity to the binning in $t$ was studied by
using an unbinned maximum likelihood method for the fit,
 which gave a result differing from the nominal one by less than 
 0.1~GeV$^2$.

\end{itemize}

The quadratic sum of these effects contributes 
$\Delta b=^{+0.5}_{-0.1}$~GeV$^{-2}$.

\end{enumerate}

\noindent 
All contributions were summed in quadrature, 
yielding a total systematic error $\Delta b=^{+1.2}_{-1.1}$~GeV$^{-2}$. 

Table~1 lists the present result together with those of 
the Fermilab photoproduction experiment E612~\cite{chapin} and 
that obtained by ZEUS for 
$\langle Q^2 \rangle=8$~GeV$^2$~\cite{lps_f2d3}. 
The present result agrees within errors with both.
This agreement suggests that at fixed $W$ 
there is little dependence of the slope on $Q^2$ and that for real 
photons the $W$ dependence is not strong. Note that
the $t$ range of our measurement is different from that of~\cite{chapin},
which is $0.02<|t|<0.1$~GeV$^2$. A direct comparison should 
therefore be made with caution: for example, in elastic $\pi p$ 
scattering~\cite{reviews}, the $t$-slope measured in the range 
$0.1 \lsim |t| \lsim 0.4$~GeV$^2$
is about 1.2~GeV$^{-2}$ lower than in the range covered by~\cite{chapin};
this difference is of the same size as the errors of our measurement. 

The weak $Q^2$ dependence of the $t$-slope in diffractive photon 
dissociation may be contrasted with the change with $Q^2$ observed 
for elastic $\rho^0$ meson 
%production~\cite{rho93}-\cite{rhodis_h1},
production~[9-13],
where $b$ decreases by approximately 3-5~GeV$^{-2}$  
when going from $Q^2\approx 0$ to $Q^2 \approx 10$~GeV$^{2}$.
If factorisation of the diffractive vertices~\cite{reviews} is assumed,
the amplitudes for the reactions $\gamma p \rightarrow X p$ and 
$\gamma p \rightarrow \rho^0 p$ are proportional to the products of 
vertex functions 
$G_{\gamma X}(Q^2,t) \cdot G_{pp}(t)$ and
$G_{\gamma \rho^0 }(Q^2,t) \cdot G_{pp}(t)$, respectively. In this framework, 
the $t$-slope includes the sum of the contributions from the $\gamma$-$X$ or
$\gamma$-$\rho^0$ vertex and from the $p$-$p$ vertex. The comparison
of the $\gamma p \rightarrow X p$  and $\gamma p \rightarrow \rho^0 p$ 
slopes indicates that the vertex function $G_{\gamma X}(Q^2,t)$ has a 
weaker dependence on $Q^2$ than $G_{\gamma \rho^0}(Q^2,t)$. A rapid $Q^2$
dependence of $G_{\gamma \rho^0}(Q^2,t)$ is expected in pQCD inspired 
models of elastic vector meson production~\cite{futurephysicsathera}, 
reflecting the decrease with $Q^2$ of the transverse size of the 
quark-antiquark pair into which the photon fluctuates before 
interacting with the proton. A weak $Q^2$ dependence of 
the function $G_{\gamma X}(Q^2,t)$ is also expected in the framework of
various models of diffractive dissociation of photons~\cite{kolya}.

{\footnotesize
\begin{table}
\begin{center}
\begin{tabular}{|l|c|c|c|} \hline
                               & Present result   & Ref.~\protect\cite{chapin} & Ref.~\protect\cite{lps_f2d3} \\ \hline
$\langle Q^2 \rangle$/GeV$^2$  & $\approx 0$      & 0                          &      8                       \\
$W$ range/GeV                  &  176-225         & 12-17                      &   50-270                     \\
$M_X$ range/GeV                &  4-32            & 1.4-1.7, 1.7-2.2, 2.2-3    &   2-27                       \\
$|t|$ range/GeV$^2$            &  0.073-0.4       & 0.02-0.1                   &   0.073-0.4                  \\ \hline
$b$/GeV$^{-2}$                 &$6.8 \pm 0.9$~(stat.)~$^{+1.2}_{-1.1}$~(syst.)&$4.2\pm 1.4$, $6.3 \pm 1.3$, $5.1 \pm 1.3$&$7.2 \pm 1.1$~(stat.)~$^{+0.7}_{-0.9}$~(syst.)\\
\hline
\end{tabular}
\end{center}
\caption{A compilation of results for the $t$-slope for the reaction 
$\gamma p \rightarrow X p$. The present result is listed together with those from 
ref.~\protect\cite{chapin} for real photons and that of 
ref.~\protect\cite{lps_f2d3} for virtual photons.
}
\label{table_slopes}
\end{table}
}

\section*{Acknowledgements}

We thank the DESY Directorate for their strong support and encouragement,
and the HERA machine group for their diligent efforts. 
Collaboration with the HERA group was particularly crucial to the successful 
installation and operation of the LPS. We are also  grateful for
the support of the DESY computing and network services.
The design, construction,
and installation of the ZEUS detector have been made possible by the ingenuity
and dedicated effort of many people from DESY and the home institutes
who are not listed as authors; among them we would like to thank B. Hubbard 
for his invaluable contributions to the experiment, and the LPS in particular.
Finally, it is a pleasure to thank N.N.~Nikolaev, M.G.~Ryskin and M.~Strikman
for many useful discussions.

\newpage

\begin{thebibliography}{99}

\bibitem{chapin} E612 Collab., T.J. Chapin et al., Phys.~Rev. {\bf D31} (1985) 17.
\bibitem{h1_mx} H1 Collab., C. Adloff et al., Z. Phys. {\bf C74} (1997) 221.
\bibitem{zeus_mx} ZEUS Collab.,  J. Breitweg et al., 
Z. Phys. {\bf C75} (1997) 421.


\bibitem{sakurai} J.J. Sakurai, Ann. Phys.  {\bf 11} (1960) 1.
\bibitem{bauer} T.H. Bauer et al., Rev. Mod. Phys. {\bf 50} (1978) 261.
\bibitem{collins} 
T. Regge, Nuovo Cimento {\bf 14} (1959) 951;\\
T. Regge, Nuovo Cimento {\bf 18} (1960) 947;\\
see also e.g. P.D.B. Collins, ``An Introduction to Regge Theory and 
High Energy Physics", 
Cambridge University Press, Cambridge~(1977).

\bibitem{field_fox} R.D. Field and G.C. Fox, Nucl. Phys. {\bf B80} (1974) 367.

\bibitem{lps_f2d3}  
ZEUS Collab., J.~Breitweg et al., DESY Report DESY 97-184, to appear 
in Z. Phys.

\bibitem{rho93} ZEUS Collab., M. Derrick et al., Z. Phys. {\bf C69} (1995) 39.
\bibitem{rho_h1} H1 Collab., S. Aid et al., Nucl. Phys. {\bf B463} (1996) 3.
\bibitem{lps_rho} ZEUS Collab., M. Derrick et al., Z. Phys. {\bf C73} 
(1997) 253.
\bibitem{rhodis_zeus} ZEUS Collab., M. Derrick et al., Phys. Lett. {\bf B356} (1995) 601.  
\bibitem{rhodis_h1} H1 Collab., S. Aid et al., Nucl. Phys. {\bf B468} (1996) 3. 
\bibitem{omega_zeus}ZEUS Collab., M. Derrick et al., Z. Phys. {\bf C73} 
(1996) 73.
\bibitem{phiphp_zeus}ZEUS Collab., M. Derrick et al., Phys. Lett. {\bf B377} 
(1996) 259.
\bibitem{phidis_h1}H1 Collab., C. Adloff et al., Z. Phys. {\bf C75} 
(1997) 607.
\bibitem{psi_h1} H1 Collab.,  S. Aid et al., Nucl. Phys. {\bf B472} (1996) 3.
\bibitem{psi_zeus}ZEUS Collab., J.Breitweg et al., Z. Phys. {\bf C75} 
(1997) 215.

\bibitem{detector_a}
ZEUS Collab., M.~Derrick et al., The ZEUS Detector, Status Report 1993, 
DESY (1993).
\bibitem{detector_b}ZEUS Collab., M.~Derrick et al., Phys.~Lett. {\bf B293} (1992) 465.

\bibitem{vxd}C.~Alvisi et~al., Nucl.~Instr. Meth. {\bf A305} (1991) 30.

\bibitem{ctd} N. Harnew et al., Nucl.~Instr.~Meth.~{\bf A279} (1989) 290;\\
B. Foster et al., Nucl.~Phys.~(Proc.~Suppl.)~{\bf B32} (1993) 181;\\
B. Foster et al., Nucl.~Instr.~Meth.~{\bf A338} (1994) 254.


\bibitem{CAL} M. Derrick et al., Nucl.~Instr.~Meth.~{\bf A309} (1991) 77;\\
A. Andresen et al., Nucl.~Instr.~Meth.~{\bf A309} (1991) 101;\\
A. Bernstein et al., Nucl.~Instr.~Meth.~{\bf A336} (1993) 23;\\
A. Caldwell et al., Nucl.~Instr.~Meth.~{\bf A321} (1992) 356.
                                          
\bibitem{lumi}
D. Kisielewska et al., DESY-HERA Report 85-25 (1985);\\
J.~Andruszk\'ow et al., DESY report DESY~92-066 (1992);\\
K. Piotrzkowski, Ph.D. Thesis, Cracow INP-Exp., DESY Internal Report 
F35D-93-06 (1993).

\bibitem{michal} M. Kasprzak, Ph.D. thesis, Warsaw University,
DESY Internal Report F35D-96-16 (1996).

\bibitem{ZEUS_sigtot}
ZEUS Collab., M.~Derrick et al., Z.~Phys.~{\bf C63}~(1994)~391.

\bibitem{Burow}
B.D. Burow, Ph.D. Thesis, University of Toronto, DESY Internal Report 
F35D-94-01 (1994).


\bibitem{ada}
P.~Bruni et al., Proc. Workshop on Physics at HERA, DESY, Eds. W. Buchm\"uller
and G.~Ingelman (1991)~363;\\
A.~Solano, Tesi di Dottorato, University of Torino (1993),
unpublished (in Italian).
\bibitem{nik_zak}
N.N.~Nikolaev and B.G.~Zakharov, Z.~Phys.~{\bf C53}~(1992)~331.

\bibitem{geant}
GEANT 3.13, R. Brun et al., CERN DD/EE/84-1 (1987).
           

\bibitem{thesis_roberto}
R. Sacchi, Tesi di Dottorato, University of Torino (1996),
unpublished (in Italian).



\bibitem{reviews} See e.g.:\\
G. Alberi and G. Goggi, Phys. Rep. {\bf74} (1981) 1;\\
K. Goulianos, Phys. Rep. {\bf101} (1983) 169;\\
M. Kamran, Phys. Rep. {\bf108} (1984) 275;\\
N.P. Zotov and V.A. Tsarev, Sov. Phys. Uspekhi {\bf 31} (1988) 119;\\
G. Giacomelli, Int. J. Mod. Phys. A, vol. 5, no. 2 (1990), 223.
 
\bibitem{futurephysicsathera}See e.g.:\\
W. Koepf et al., in Proceedings of the ``Workshop on Future Physics at 
HERA", Hamburg, Sept. 1995-May 1996, Editors G. Ingelman, A. De Roeck, 
R. Klanner, p.~674, and references therein;\\
ibid. H. Abramowicz et al., p.~679, and references therein.

\bibitem{kolya}See e.g.:\\
H.~Abramowicz et al., in Proceedings of 
the ``22nd Annual SLAC Summer Institute on Particle Physics", Stanford,
California, 8-19 Aug. 1994, Editors L.~De Porcel and J.~Chan, SLAC, 1996,
p.~539, DESY Report DESY 95-047, and
M.~Strikman, private communication;\\ 
M.G. Ryskin, J. Phys. {\bf G22} (1996) 741, and private communication;\\
N.N.~Nikolaev and B.G. Zakharov, preprint hep-ph/9706343~v2, and 
Proceedings of the ``Fifth International Workshop on Deep Inelastic 
Scattering and QCD", Chicago, April 14-18, 1997, Editors D. Krakauer and 
J. Repond, American Institute of Physics Conference Proceedings no. 407, 
p.~445, and references therein.

\end{thebibliography}
\end{document}